\documentclass[prb,twocolumn,showpacs,floatfix]{revtex4}
\usepackage{graphicx}
\usepackage{amssymb}
\usepackage{amsmath}
\usepackage{xspace}
\usepackage{url}

\begin{document} 
\title{Phonon-mediated sticking of electrons at dielectric surfaces}

\author{R. L. Heinisch, F. X. Bronold, and 
H. Fehske}
\affiliation{$^{1}$Institut f{\"ur} Physik,
             Ernst-Moritz-Arndt-Universit{\"a}t Greifswald,
             17489 Greifswald,
             Germany}

\date{\today}
\begin{abstract}
We study phonon-mediated temporary trapping of an electron in polarization-induced 
external surface states (image states) of a dielectric surface. Our approach is based 
on a quantum-kinetic equation for the occupancy of the image states. It allows us to 
distinguish between prompt and kinetic sticking. Because the depth of the image potential 
is much larger than the Debye energy multi-phonon processes are important. Taking two-phonon
processes into account in cases where one-phonon processes yield a vanishing transition 
probability, as it is applicable, for instance, to graphite, we analyze the adsorption
scenario as a function of potential depth and surface temperature and calculate prompt 
and kinetic sticking coefficients. We find rather small sticking coefficients, at most 
of the order of $10^{-3}$, and a significant suppression of the kinetic sticking coefficient 
due to a relaxation bottleneck inhibiting thermalization of the electron with the surface 
at short timescales. 
\end{abstract}
\maketitle

\section{Introduction}

A complete kinetic modeling of atmospheric~\cite{RL01}, interstellar~\cite{Whipple81,DS87,Mann08}, 
or man-made bounded plasmas~\cite{FIK05,Ishihara07,GMB02,Kogelschatz03,KCO04,SAB06,SLP07,LLZ08}
requires boundary conditions for the distribution functions of the relevant plasma species 
(electrons, ions, neutrals), that is, a quantitative microscopic understanding of the 
elementary processes at the plasma boundary. Of particular importance is the build-up of 
a quasi-stationary negative surface charge, which not only depletes the electron
density in front of the boundary (sheath formation) but also acts as an electron reservoir
for surface-supported electron-ion recombination and secondary electron emission which in 
turn affect the charge balance in the bulk of the plasma.~\cite{LL05} Despite its 
unquestioned importance, little is quantitatively known about the microphysics of electrons
at plasma boundaries. It is only until recently that 
we proposed that the charging of plasma boundaries can be perhaps microscopically understood
in terms of an electronic physisorption process.~\cite{BFKD08,BDF09}

The physisorption scenario applies to a plasma electron approaching a metallic  or a dielectric  
boundary provided its kinetic energy is large enough to overcome the Coulomb barrier due to the 
charges already residing on the boundary but small enough to make the surface appear as having a 
negative electron affinity. If the electron can convert its energy into internal 
energy of the boundary, via exciting elementary excitations of the solid, it may get stuck 
(adsorbed) at the boundary. Later it may desorb again if it gains enough energy from the boundary. 

Like physisorption of neutral
particles~\cite{BY73,IN76,GKT80a,GKT80b,Leuthaeusser81,Brenig82,KG86,CG88,GS91,AM91,CK92,BG93,BGB93,BGR94,CC98}
physisorption of electrons is the polarization-induced temporary binding to a surface. It can be 
characterized by a desorption time and a sticking coefficient. At first 
glance physisorption of electrons seems to be not much different from physisorption of 
neutral particles. There are however important qualitative differences which warrant a separate 
theoretical investigation.

First, albeit not in the focus of our investigation, the long-range $1/z$-tail 
of the image potential leads to a finite electron sticking coefficient at vanishing electron energy
and surface temperature.~\cite{CK92,BR92} This is in contrast to the quantum reflection, that is, 
the vanishing sticking coefficient, one finds in this limit for 
the short-ranged surface potentials typical for physisorption of neutral particles.~\cite{CG88,CC98} 

Second, the surface potential in which physisorption of electrons occurs, in particular at plasma 
boundaries, consists of a polarization-induced attractive part and a repulsive Coulomb part due to 
electrons already adsorbed on the surface. The limit of vanishing coverage, very often adopted in 
the theoretical description of physisorption of neutral particles, is thus only applicable to the
very first (last) electron approaching (leaving) the boundary.

Third, in contrast to physisorption of neutral particles, physisorption of electrons
has to be always described quantum mechanically because the image potential varies on
a scale comparable to the thermal de Broglie wave length of the electron.~\cite{BFKD08}
This is also the case for physisorption of positronium.~\cite{NNS86,MSL91,WJS92}

Finally, and this will be the theme of our investigation, the polarization-induced image 
potential supports deep states, in addition to shallow ones. Direct transitions 
from the continuum to deep bound states are very unlikely. Hence, a modeling in terms of a 
quantum-kinetic rate equation for the occupancy of the bound surface states,~\cite{Brenig82,KG86} 
and Brenig's distinction between prompt and kinetic sticking,~\cite{Brenig82} 
is vital for a correct description of electron physisorption. For phonon-controlled adsorption
and desorption, as it occurs at dielectric surfaces, deep states also imply that multi-phonon processes 
have to be taken into account in the calculation of state-to-state transition probabilities. This can be 
done either via an expansion of the energy dependent $T$-matrix~\cite{BY73,GKT80a,AM91}, the method we are 
using~\cite{HBF10}, or via a Magnus-type resummation of the time-dependent scattering 
operator.~\cite{Gumhalter96,Gumhalter01,SG03,SG05,SG08}

In the following we investigate adsorption of an electron to a dielectric surface at finite 
temperature assuming an acoustic longitudinal bulk phonon controlling electron energy relaxation 
at the surface. To avoid complications due to finite coverage we focus on the first approaching
electron. Using the quantum-kinetic rate equation for 
the occupancy of the image states of our previous work~\cite{HBF10} (thereafter referred 
to as I), where we studied desorption of an image-bound electron from a dielectric surface, 
we calculate prompt and kinetic sticking coefficients. Compared to semiclassical 
estimates~\cite{UN80} they turn out to be extremely small. Instead of the order of $10^{-1}$ 
we find them to be at most of the order of $10^{-3}$. We also analyze in detail the adsorption 
scenario as a function of surface temperature and potential depth. Most notable, our results
reveal an energy relaxation bottleneck prohibiting, on a short timescale, thermalization 
of the electron with the surface, that is, the trickling through of the electron from upper 
to deep states. The reduced accessibility of deep states makes the kinetic sticking 
coefficient much smaller than the prompt sticking coefficient in contrast to what is usually found in 
physisorption of neutral particles.~\cite{BGB93}

The remaining paper is structured as follows. In Sec. II, we specify the quantum-kinetic approach
of our previous work concerning desorption (paper I~\cite{HBF10}) to the situation of adsorption 
and introduce prompt and kinetic sticking coefficients. We then briefly recall in Sec. III  
the calculation of the state-to-state transition probabilities based on a microscopic model 
for the electron-surface interaction and an expansion of the $T$-matrix for the dynamic part of 
that interaction. Mathematical details not given can be found in I. 
Finally, in Sec. IV, we present and discuss our results before we conclude in Sec. V.

\section{Electron kinetics}

The probability for an approaching electron in the continuum state \(k\) to make a transition to any of the bound 
states \(n\) of the polarization-induced image potential is given by the prompt energy-resolved sticking 
coefficient,~\cite{KG86}
\begin{align}
s_{e,k}^\text{prompt} = \tau_t \sum_n W_{nk} \text{ ,}
\end{align}
where \(\tau_t=2L/v_z\) is the travelling time through the surface potential of width \(L\) which, in the 
limit \(L\rightarrow \infty\), can be absorbed into the transition probability from the continuum 
state \(k\) to the bound state \(n\), \(W_{nk}\). If the incident unit electron flux (we consider only 
a single electron impinging on the surface) is stationary and corresponds to an electron with 
Boltzmann distributed kinetic energies, the prompt energy-averaged sticking coefficient is 
given by~\cite{KG86}
\begin{align}
s^{\rm prompt}_e=\frac{\sum_k s^{\rm prompt}_{e,k} k 
e^{-\beta_e E_k}}{\sum_k k e^{-\beta_e E_k}} \text{ ,} \label{promptenergyavergaed}
\end{align}
where \(\beta_e^{-1}=k_BT_e\) is the mean electron energy.

Prompt sticking coefficients are properly weighted sums over state-to-state transition probabilities from
continuum to bound surface states. They give the probability for initial trapping, which, according to Iche 
and Nozi\'eres\cite{IN76} and Brenig~\cite{Brenig82}, is the first out of three stages of physisorption. The 
second stage encompasses relaxation of the bound state occupancy and the third stage is the 
desorption of the temporarily bound particle. 

The second stage, which also includes transitions back to the continuum, cannot be captured by simple 
state-to-state transition probabilities. Instead, a quantum-kinetic rate equation for the time-dependent 
occupancy of the bound surface states \(n_n(t)\) has to be employed.~\cite{GKT80a,Brenig82} This 
equation describes processes on a timescale much longer than the lifetime of the individual surface 
states but shorter than the desorption time.~\cite{GKT80a,Brenig82,KG86} Following Gortel and 
coworkers,~\cite{GKT80a,KG86} 
\begin{align}
\frac{\mathrm{d}}{\mathrm{d}t}n_n(t)=&\sum_{n^\prime} \left[W_{n n^\prime} n_{n^\prime}(t) - W_{n^\prime n} 
n_n(t) \right] \nonumber \\
& -\sum_k W_{k n} n_n(t) +\sum_k \tau_t W_{nk} j_k(t) \text{ ,} \label{fullrateeqn}
\end{align}
where \(W_{n^\prime n}\) is the probability for a transition from a bound state \(n\) to another bound state 
\(n^\prime\), \(W_{kn}\) and \(W_{nk}\) are the probabilities, respectively, for a transition from a bound 
state \(n\) to a continuum state \(k\) and vice versa, and 
\begin{eqnarray}
j_k(t)=n_k(t)\tau^{-1}_t  
\end{eqnarray}
is the incident electron flux which in principle can be non-stationary.

The solution to Eq. (\ref{fullrateeqn}) can be obtained from the solution of the corresponding 
homogeneous equation,
\begin{align}
\frac{\mathrm{d}}{\mathrm{d}t}n_n(t)&=\sum_{n^\prime} \left[W_{n n^\prime} n_{n^\prime}(t) 
- W_{n^\prime n} n_n(t) \right]  -\sum_k W_{k n} n_n(t)\nonumber\\
&= T_{nn^\prime}n_{n^\prime}(t)~,  \label{homorateeqn}
\end{align}
and treating the electron flux \( j_k(t)\) as an externally specified quantity.~\cite{Brenig82,KG86}
In the simplest case, which is also the basis of Eq.~(\ref{promptenergyavergaed}), 
\( j_k(t)\) is the stationary flux corresponding to a single electron whose energy 
is Boltzmann distributed over the continuum states \(k\) with a mean electron energy \(k_BT_e\), 
that is, \(j_k(t)\equiv j_k\sim k e^{-\beta_eE_k}\). 

To solve Eq. (\ref{homorateeqn}) amounts to solving the eigenvalue problem for the matrix 
${\bf T}$.~\cite{Brenig82,KG86} For the particular case of an electron physisorbing at a dielectric
surface this has been already done in I. If the transitions between bound states are much faster than 
the transitions to the continuum, so that the adsorbed electron escapes very slowly, one eigenvalue, 
\(-\lambda_0\), turns out to be considerably smaller than all the other eigenvalues \(-\lambda_\kappa\). 
The equilibrium occupation of the bound states, \(n_n^\mathrm{eq}\), is then to a very good approximation 
the right eigenvector to \(-\lambda_0\), which can be thus identified with the negative of the inverse of 
the desorption time, that is, \(\lambda_0=\tau_e^{-1}\).

The kinetic sticking coefficient, which takes into account not only the initial capture but also the subsequent 
relaxation of the occupancy of the bound surface states, can be obtained as follows.~\cite{KG86} The 
solution of Eq.~(\ref{fullrateeqn}) is split according to 
\begin{align}
n_n(t)=n_n^{\rm s}(t)+n_n^{\rm f}(t)~, 
\label{ntotal}
\end{align}
where
\begin{align}
n_n^{\rm s}(t)=e^{-\lambda_0 t} \int_{-\infty}^t \mathrm{d}t^\prime e^{\lambda_0 t^\prime} e_n^{(0)} 
\sum_{k,l} \tilde{e}_l^{(0)} \tau_t W_{lk} j_k(t^\prime) \text{  }
\label{nslow}
\end{align}
is the slowly and
\begin{align}
n_n^{\rm f}(t)=\!\sum_{\kappa \neq 0} e^{-\lambda_\kappa t}\!\!\int_{-\infty}^t\!\!\!\!\!\mathrm{d}t^\prime e^{\lambda_\kappa t^\prime} 
e_n^{(\kappa)} \sum_{k,l} \tilde{e}_l^{(\kappa)} \tau_t W_{lk} j_k(t^\prime) \text{ }
\label{nquick}
\end{align}
the quickly varying part of \(n_n(t)\). The quantities \(e_n^{(\kappa)} \) and 
\(\tilde{e}_n^{(\kappa)}\) are, respectively, the 
components of the right and left eigenvectors of the matrix \({\bf T}\) to the eigenvalue \(-\lambda_\kappa\). 
The probability of the electron remaining in the surface states for times of the order of the desorption time is 
given by the slowly varying part only, that is, \(n^{\rm s}(t)=\sum_n n_n^{\rm s}(t)\). Differentiating 
\(n^{\rm s}(t)\) with respect to time,
\begin{align}
\frac{d}{dt}n^{\rm s}(t)=\sum_k s^{\rm kinetic}_{e,k} j_k(t) -\lambda_0 n^{\rm s}(t)~,
\end{align}
enables us, following Brenig~\cite{Brenig82}, to identify the kinetic energy-resolved sticking coefficient,
\begin{align}
s_{e,k}^\text{kinetic}=\tau_t \sum_{n,l} e_{n}^{(0)} \tilde{e}_l^{(0)} W_{lk} \text{ ,}
\end{align}
which gives the probability for the electron being trapped even after the energy relaxation of the second
stage of physisorption. In analogy to Eq. (\ref{promptenergyavergaed})
the energy-averaged kinetic sticking coefficient reads for a stationary Boltzmannian electron flux 
\begin{align}
s^{\rm kinetic}_e=\frac{\sum_k s^{\rm kinetic}_{e,k} k e^{-\beta_e E_k}}{\sum_k k e^{-\beta_e E_k}} \text{ .} 
\label{kineticenergyavergaed}
\end{align}

\section{Transition probabilities}
\label{Electron-surface interaction and transition probabilities}

The transition probabilities \(W_{qq^\prime}\), where \(q\) and \(q^\prime\) stand either for \(k\) or \( n\), 
are the fundamental building blocks of the foregoing analysis. They have to be calculated from a microscopic
model for the electron-surface interaction. The necessary steps have been described in I. 

In short, for a dielectric surface, the main source, leaving interband electronic excitations aside, 
which primarily affect the dielectric constant, of the attractive static electron-surface potential 
is the coupling of the electron to a dipole-active surface phonon.~\cite{RM72,EM73} Far from the 
surface the surface potential merges with the classical image potential and thus \(\sim 1/z\). Close 
to the surface, however, it is strongly modified by the recoil energy resulting from the momentum 
transfer parallel to the surface when the electron absorbs or emits a surface phonon. Taking this 
effect into account leads to a recoil-corrected image potential \(\sim 1/(z+z_c)\) with $z_c$ a 
cut-off parameter defined in I. Transitions between the surface states supported by the image 
potential are caused by a longitudinal acoustic bulk phonon perpendicular to the surface. 

The Hamiltonian from which we calculated the transition probabilities was introduced in I where 
all quantities entering the Hamiltonian are explicitly defined. It is given by~\cite{HBF10}
\begin{align}
H=H_e^\text{static} +H_{ph}+H_{e-ph}^\text{dyn} \text{ ,} 
\label{Htotal}
\end{align}
where
\begin{align}
H_e^\text{static}=\sum_q E_q c_q^\dagger c_q 
\end{align}
describes the electron in the recoil-corrected image potential, which thus accounts for the coupling 
of the electron to the dipole-active surface phonon, 
\begin{align}
H_{ph}=\sum_Q \hbar \omega_Q b_Q^\dagger b_Q \text{ ,}
\end{align}
describes the free dynamics of the acoustic bulk phonon responsible for transitions between surface 
states, and 
\begin{align}
H_{e-ph}^\text{dyn}=\sum_{q,q^\prime} \langle q^\prime | V_p(u,z)|q\rangle c_{q^\prime}^\dagger c_q \text{ .}
\end{align}
denotes the dynamic coupling of the electron to the bulk phonon. Expanding $V_p(u,z)$ with respect to 
the displacement field,
\begin{align}
u=\sum_Q \sqrt{\frac{\hbar}{2\mu \omega_Q N_s}} \left( b_Q+b_{-Q}^\dagger \right) \label{uomegarel} \text{ ,}
\end{align}
allows us to classify the dynamic interaction according to the number of exchanged bulk phonons.

As in I we use a Debye model for the bulk phonon. Sums over phonon momenta are thus replaced by
\begin{align}
\sum_Q \dots=\frac{3N_s}{\omega_D^3}\int \mathrm{d}\omega \omega^2 \dots  \text{ .} \label{debyemodel}
\end{align}
Measuring energies in units of the Debye energy \(\hbar \omega_D=k_BT_D\), important dimensionless energy
parameters characterizing~(\ref{Htotal}) are  
\begin{align}
\epsilon_n=\frac{E_n}{\hbar \omega_D} \quad \text{and} 
\quad \Delta_{nn^\prime}=\frac{E_n-E_{n^\prime}}{\hbar \omega_D} \text{ ,}
\end{align}
where \(E_n < 0\) is the energy of the \(n^\text{th}\) 
bound state. We call the surface potential shallow if the lowest bound state is at most one Debye energy 
beneath the continuum, that is, \(\epsilon_1>-1\), one-phonon deep if the energy difference between the 
lowest two bound states does not exceed one Debye energy, that is, \(\Delta_{12}>-1\), two-phonon deep 
if the energy difference between the lowest two bound states is between one and two Debye energies, that is, 
\(-1>\Delta_{12}>-2\), and so forth.

\begin{table}
\caption{Dielectric constant \(\epsilon_s\), Debye energy \(\hbar\omega_D\), energy difference of the lowest 
two bound states of the recoil-corrected image potential \(\Delta E_{12}\), and the corresponding potential 
depth parameter \(\Delta_{12}\) for graphite, silicon dioxide,
and gallium arsenide. }
\center
\begin{tabular}{c c c c c c}
 &\(\epsilon_s\) & \(\hbar\omega_D\) & \(\Delta E_{12} \) & \(\Delta_{12}\) \\
\hline 
graphite      & \(13.5\) & \(0.215\)eV & \(0.233\)eV & \(1.06\) \\
${\rm SiO_2}$ & \(3.8\)  & \(0.041\)eV & \(0.105\)eV & \(2.59\) \\
GaAs          & \(13\)   & \(0.030\)eV & \(0.152\)eV & \(5.13\)
\end{tabular}
\label{materialcomp}
\end{table}

Because of the strong interaction between the electron and the dipole-active surface phonon, physisorption 
of an electron typically takes place in an at least two-phonon deep image potential (see Table \ref{materialcomp}). 
Hence, physisorption of an electron controlled by a bulk acoustic phonon, as anticipated in (\ref{Htotal}) 
and in fact applicable to dielectric surfaces, where large energy gaps block electronic relaxation channels 
due to internal electron-hole pairs and/or plasmons, has to involve the exchange of many bulk phonons.

The transition probability from an electronic state \(q\) to an electronic state \(q^\prime\) is given by\cite{BY73}
\begin{align}
\mathcal{R}(q^\prime,q)=&\frac{2\pi}{\hbar} \sum_{s,s^\prime} 
\frac{e^{-\beta_s E_s}}{\sum_{s^{\prime\prime}}e^{-\beta E_{s^{\prime\prime}}}} 
|\langle s^\prime, q^\prime |T|s,q\rangle|^2 \nonumber \\
&\times \delta(E_s-E_{s^{\prime}}+E_q-E_{q^\prime}) \text{ ,}
\end{align}
where $T$ is the on-shell $T$-matrix corresponding to \(H_{e-ph}^\text{dyn}\) and \(\beta_s=(k_BT_s)^{-1}\) with \(T_s\) the 
surface temperature; \(|s\rangle\) and \(|s^\prime \rangle\) are initial and final phonon states, which are 
averaged over. 

Multi-phonon processes have two possible origins.~\cite{BY73} They arise from the expansion of 
\(H_{e-ph}^\text{dyn}\) with respect to $u$, 
\begin{align}
H_{e-ph}^\text{dyn}=V_1+V_2+V_3+\mathcal{O}(u^4)
\end{align}
and from the multiple action of this perturbation. Defining the free electron-phonon resolvent, 
\begin{align}
G_0=(E-H_e^\text{static}-H_{ph}+i\epsilon)^{-1} \text{ ,}
\end{align}
the latter is encoded in the $T$-matrix equation,
\begin{align}
T=H_{e-ph}^\text{dyn}+H_{e-ph}^\text{dyn}G_0T \text{ .}
\end{align}

Using the short-hand notation introduced in I, the one-phonon process, proportional to \(u^2\), is 
accounted for by
\begin{align}
\langle s^\prime,q^\prime | V_1 |s,q\rangle \langle s,q|V_1^\ast | s^\prime, q^\prime \rangle~.
\end{align}
It leads to the standard golden rule approximation for the transition probability. 

Two-phonon processes are proportional to \(u^4\) and thus less likely than one-phonon processes. Most of 
them renormalize only the one-phonon transition probability and can thus be neglected in a first approximation. 
There are however two-phonon processes which induce transitions absent in the one-phonon approximation and hence
have to be included in the calculation of the transition probabilities. In our short-hand notation the processes 
in question are 
\begin{align}
&\langle s^\prime,q^\prime | V_2 |s,q\rangle \langle s,q|V_2^\ast | s^\prime, q^\prime \rangle~, \label{v2square} \\
&\langle s^\prime,q^\prime | V_2 |s,q\rangle \langle s,q|V_1^\ast G_0^\ast V_1^\ast | s^\prime, q^\prime \rangle~,  \label{v2v1square} \\
&\langle s^\prime,q^\prime | V_1 G_0 V_1 |s,q\rangle \langle s,q|V_2^\ast | s^\prime, q^\prime \rangle~, \label{v1squarev2} \\
&\langle s^\prime,q^\prime | V_1 G_0 V_1 |s,q\rangle \langle s,q| V_1^\ast G_0^\ast V_1^\ast | s^\prime, q^\prime \rangle~. \label{v1quartic} 
\end{align}
It is shown in I how these processes can be included in the calculation of the transition probabilities
\(W_{qq^\prime}\) entering the rate equation~(\ref{fullrateeqn}). Singularities appearing in some of 
the two-phonon transition rates have been regularized by taking a finite phonon lifetime into account 
(see I and Ref.~\cite{Rafael09} for details).

The electronic matrix elements entering the transition probabilities have been also calculated in I, 
using bound and unbound wavefunctions of the recoil-corrected image potential. Hence, not only
bound states but also continuum states belong to the static surface potential.~\cite{HBF10} Our 
approach is thus on par with the distorted-wave Born approximation employed, for instance, by Armand and 
Manson for the calculation the sticking coefficient for light neutral particles.~\cite{AM91} 

\section{Results}

The material parameters chosen for the numerical calculations are, unless specified otherwise, given 
in Table~\ref{materialtable}. They correspond to graphite. For some calculations we use however
the Debye temperature as a tunable parameter to realize different potential depths which is the main focus 
of this investigation.

\begin{table}
\caption{Material parameters for the numerical results.   }
\center
\begin{tabular}{l l l l l}
\hline 
Debye temperature & \(\quad\) & \(T_D\) & \(\quad\) & \(2500\)K \\
Dielectric constant & \(\quad\) & \(\epsilon_s\) & \(\quad\) & \(13.5\) \\ 
TO phonon mode frequency & \(\quad\) & \(\hbar \omega_T\) & \(\quad\) & \(170 \text{meV}\)  \\
Gr\"uneisen parameter & \(\quad\) &  \(\gamma_G\) & \(\quad\) & \(1.7\) \\
Shear modulus & \(\quad\) & \(\mu  \) & \(\quad\) & \(5\) GPa  \\
\hline 
\end{tabular}
\label{materialtable}
\end{table}

\subsection{One-phonon deep potentials}
 
First, we present results for shallow and one-phonon deep surface potentials.
In leading order, only one-phonon processes are involved and the one-phonon 
approximation for the transition probabilities is sufficient. Because the electron thermalizes 
then very quickly with the surface the prompt and kinetic sticking coefficients are almost 
identical. In this subsection we show therefore only results for the prompt sticking coefficient. 

\begin{figure}
\begin{minipage}{0.49\linewidth}
\includegraphics[width=0.8\linewidth]{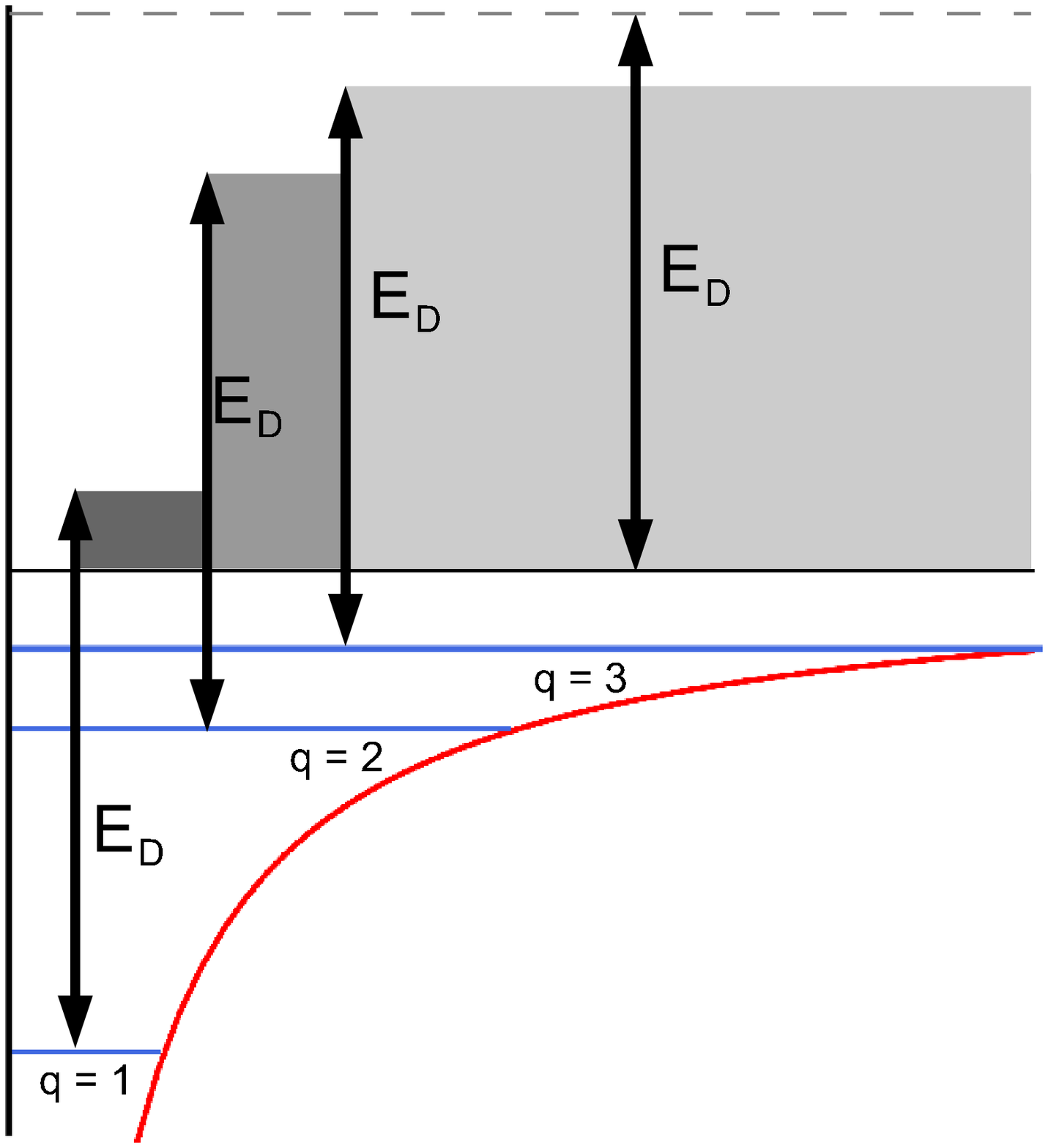}
\end{minipage}
\begin{minipage}{0.49\linewidth}
\includegraphics[width=0.8\linewidth]{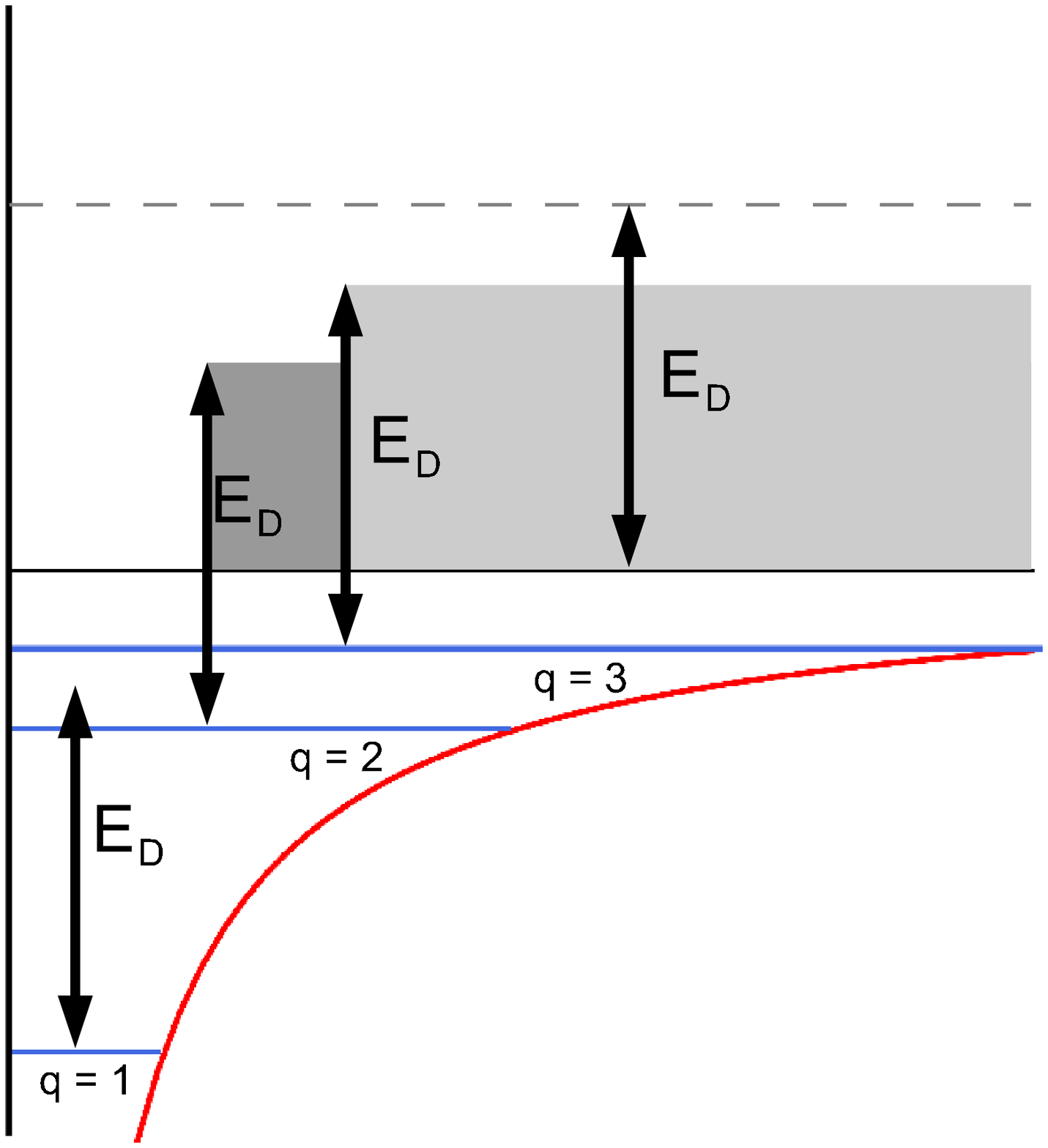}
\end{minipage}

\begin{minipage}{0.49\linewidth}
\includegraphics[width=\linewidth]{figure1b.eps}
\end{minipage}
\begin{minipage}{0.49\linewidth}
\includegraphics[width=\linewidth]{figure1e.eps}
\end{minipage}

\begin{minipage}{0.49\linewidth}
\includegraphics[width=\linewidth]{figure1c.eps}
\end{minipage}
\begin{minipage}{0.49\linewidth}
\includegraphics[width=\linewidth]{figure1f.eps}
\end{minipage}
\caption{Upper panel: Sketch of a shallow (left) and one-phonon deep (right) potential. The grey shaded
areas show the energy range of sticking by one-phonon processes.
Middle panel: Energy-resolved prompt sticking coefficient as a function of the electron energy for a
shallow potential (\(T_D=4100{\rm K}\)) at \(T_s=410{\rm K}\) (left) and for a one-phonon deep potential
(\(T_D=3000{\rm K}\)) at \(T_s=300{\rm K}\) (right).
Lower panel: Energy-averaged prompt sticking coefficient as a function of the mean electron energy
for a shallow potential (\(T_D=4100{\rm K}\)) at \(T_s=205{\rm K}\) (left) and a one-phonon deep potential 
(\(T_D=3000{\rm K}\))
at \(T_s=150{\rm K}\) (right).
}
\label{figure1}
\end{figure}

Figure \ref{figure1} compares \(s^{\rm prompt}_e\) for a shallow and a one-phonon deep potential. The sketches 
in the upper part of the figure illustrate the main difference between the two potentials. For a 
shallow potential the lowest bound state is less than one Debye energy below the continuum so that a low-lying 
electron from the continuum can be directly trapped in the lowest bound state, \(n=1\), by a one-phonon transition.
In the case of a one-phonon deep potential, one-phonon processes can only lead to trapping in one of the upper 
bound states \( n>1\). 

The middle panels of Fig.~\ref{figure1} show the prompt energy-resolved sticking coefficient as a function of 
the energy of the incident electron. Apart from discontinuities the sticking coefficient depends linearly on 
the electron energy. As explained in Sec.~\ref{Electron-surface interaction and transition probabilities} the 
one-phonon transition probability is proportional to \(u^2\). From Eq.~(\ref{uomegarel}) we have \(u^2 \sim 1/\omega\) 
so that in conjunction with the Debye model~(\ref{debyemodel}) the transition probability is proportional to 
\(\omega\) which translates due to energy conservation to a proportionality to the electron energy. 
The phonon spectrum is thus reflected in the (one-phonon) energy-resolved sticking coefficient, as it is, for 
instance, also in the cross section for (one-phonon) inelastic particle-surface scattering.~\cite{Gumhalter01}

Steep jumps in the energy-resolved sticking coefficient reflect the level accessibility. When the energy difference 
between the electron and a bound state exceeds the Debye energy, one-phonon transitions are no longer possible
and the electron can no longer directly reach that level. For a shallow potential, the first drop reflects therefore 
the accessibility of the first bound state, whereas for a one-phonon deep potential, where this bound state cannot 
be directly reached, the first drop reflects the accessibility of the second bound state. As energy differences
between successive bound states of the image potential decrease towards the ionization threshold, that is, with 
increasing \(n\) (see upper panels of Fig.~\ref{figure1}), more such steps are found near the maximum electron 
energy allowing for trapping, which is the Debye energy. 

The contribution of the \(n^\mathrm{th}\) bound state to the sticking coefficient, reflected in the height 
of the corresponding accessibility threshold, decreases for higher bound states. The reason for this lies 
in the electronic matrix element appearing in first order perturbation theory, 
\(\langle n |1/(z+z_c)^2 |k \rangle\). This matrix element is 
smaller for higher bound states because higher bound states have less weight near the surface 
where the perturbation is strongest. Of considerable importance is hence the lowest bound state, which, if
available, yields a particularly large contribution. The decreasing electronic matrix element also implies
that neglecting all but a few, say seven, of the infinitely many bound states suffices for the calculation
of the sticking coefficient. 

The prompt energy-averaged sticking coefficient is shown in the lower panels of Fig.~\ref{figure1} as 
function of the mean electron energy. Due to 
thermal averaging the sticking coefficient does no longer exhibit characteristics of the phonon spectrum 
and level accessibility, making it thus more robust against changes in the phonon model. 
It does however reflect the importance of the lowest bound state for shallow potentials. Note also, 
due to the long-range tail of the recoil-corrected image potential $\sim 1/(z+z_c)$ the energy-resolved and 
the energy-averaged electron sticking coefficients are finite for vanishing electron energy and 
electron temperature, irrespective of the surface temperature, as it should be.~\cite{BR92,CK92}

\subsection{Two-phonon deep potentials}

We now turn our attention to two-phonon deep potentials.
Under the assumption that the true one-phonon process dominates the corrections coming from two-phonon processes, 
the latter are only taken into account for transitions where one-phonon processes alone would yield no transition 
probability. 

Two-phonon processes affect in a two-phonon deep potential sticking in two ways. They enable prompt trapping from 
higher-lying continuum states, outside the one-phonon trapping range, and they control the energy relaxation 
of the trapped electron and thus the formation of the quasi-stationary 
bound state occupancy from which desorption occurs. There are thus two questions to be answered: How
significant are two-phonon processes for prompt sticking and how does the relaxation thereafter depend on
the type of phonon processes available. 

\begin{figure}
\includegraphics[width=\linewidth]{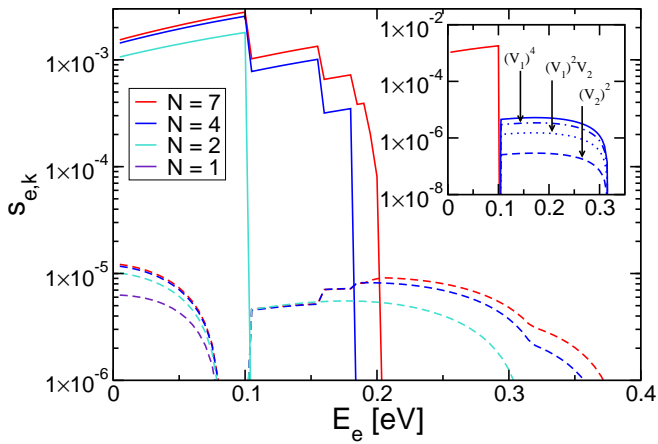}
\caption{Energy-resolved prompt sticking coefficient for a two-phonon deep potential (\(T_D=2500{\rm K}\) and 
\(T_s=500{\rm K})\) 
calculated with different numbers of bound states \(N\). Full lines denote the one-phonon contribution, dashed 
lines the two-phonon contribution. Inset: Contribution of the second bound state. One-phonon contribution (red), 
two-phonon contribution (blue) broken down into the processes \((V_2)^2\), \((V_1)^2V_2\) and \((V_1)^4\).}
\label{figure2}
\end{figure}
To address the first question we show in Fig.~\ref{figure2} the contributions to the prompt energy-resolved 
sticking coefficient arising from, respectively, one- and two-phonon processes. If available, one-phonon
processes provide for much larger sticking coefficients than two-phonon processes. Figure~\ref{figure2} 
also confirms that the sticking coefficient saturates quickly with the number of bound states included into 
the calculation. 

To investigate the relative importance of the various two-phonon processes arising, 
respectively, from the expansion of the dynamical perturbation and the $T$-matrix we plot in the 
inset of Fig.~\ref{figure2} the partial contributions to the prompt sticking coefficient arising 
from the various two-phonon processes which trigger transition to the second bound state. A 
two-phonon process can be simultaneous, as encoded in \(V_2\), or successive, as described by \(V_1G_0V_1\). 
Hence, the total two-phonon transition probability contains a contribution without virtual intermediate states, 
symbolically denoted by \(V_2^2\) [see Eq. (\ref{v2square})] and two contributions with virtual intermediate 
states, symbolically denoted by \((V_1)^2V_2\) and \((V_1)^4\) [see Eqs.  (\ref{v2v1square}), (\ref{v1squarev2}),
and  (\ref{v1quartic})]. The prompt energy-resolved sticking coefficient calculated with either \(V_2^2\),
\((V_1)^2V_2\), or \(V_1^4\) only is shown in the inset of  Fig.~\ref{figure2}. In accordance to what  
we found in our calculation of the desorption time of an image-bound electron (paper I) and to what 
Gumhalter and \v{S}iber found in their calculation of the cross section for inelastic particle-surface
scattering~\cite{SG03,SG05,SG08}, the direct two phonon process \(V_2^2\) is dominated by the processes
\(V_1^2V_2\) and \((V_1)^4\). 

Having clarified that two-phonon processes lead to a much smaller prompt sticking coefficient than one-phonon 
processes we now move on to study the effect of two-phonon transitions on the relaxation of the bound state 
occupancy. For a two-phonon deep potential the energy difference between the lowest two bound states exceeds 
one Debye energy. Hence, the relaxation of an electron trapped in one of the upper bound states  
to the quasi-stationary occupancy can only occur via two-phonon processes. Since the kinetic sticking
coefficient gives the probability for the incident electron making not only a transition to a bound state
but also a subsequent relaxation to the quasi-stationary occupancy of these states, the importance 
of two-phonon processes should be signalled by the amount the kinetic sticking coefficient deviates from 
the prompt sticking coefficient.

\begin{figure}
\includegraphics[width=\linewidth]{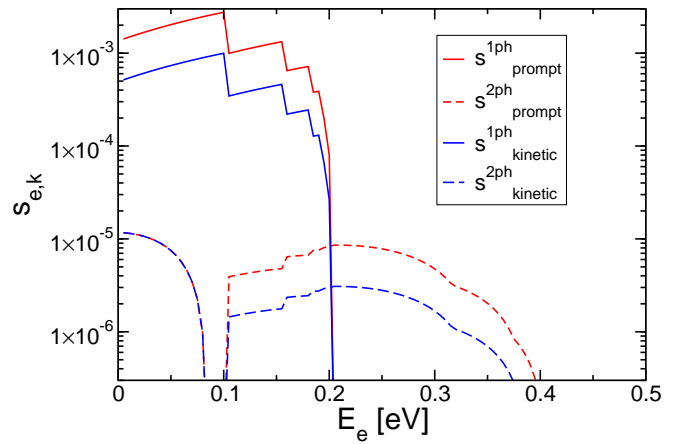}
\caption{Prompt and kinetic energy-resolved sticking coefficient as a function of the electron energy for a 
two-phonon deep potential (\(T_D=2500{\rm K}\) and \(T_s=357.14{\rm K})\). Full lines one-phonon contribution, dashed 
lines two-phonon contribution. }
\label{figure3}
\end{figure}

Figure \ref{figure3} shows that for a two-phonon deep potential the kinetic energy-resolved sticking coefficient 
is for intermediate electron energies considerably smaller than the prompt energy-resolved sticking coefficient. 
This is due to the fact that the two-phonon transition to the lowest bound state, where the major part of the 
quasi-stationary occupancy resides, is very unlikely and thus very slow. Only for small energies, the first hump 
of the two-phonon contribution to the sticking coefficient, due to trapping in the lowest bound state, are 
prompt and kinetic sticking coefficients identical because no trickling through is needed.

\begin{figure}
\includegraphics[width=\linewidth]{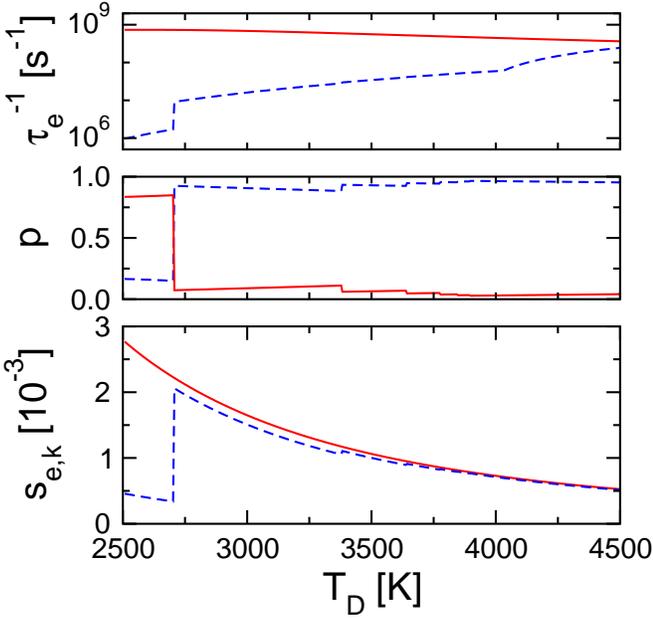}
\caption{Upper panel: Inverse desorption time from the lowest level (dashed blue line) and the upper
levels (full red line).
Middle panel: Probability for an electron initially trapped in one of the upper levels of the surface 
potential (\(n=2,3,4\dots \)) either to fall to the lowest bound state (dashed blue line) or to desorb 
without ever reaching the lowest bound state (full red line). Lower panel: Prompt (full red line) and kinetic 
(dashed blue line) energy-resolved sticking coefficient. In all three panels, \(E_e=0.1{\rm eV}\) and 
$T_s/T_D=0.2$ (to keep the level of phonon excitation constant we set $T_D/T_s$ constant~\cite{HBF10}). 
For \(T_D<2707{\rm K}\) the surface potential is two-phonon deep, 
for \(2707{\rm K} < T_D < 4029{\rm K} \) it is one-phonon deep, and for \(T_D > 4029{\rm K}\) it is shallow.} 
\label{figure4}
\end{figure}

\begin{figure}
\includegraphics[width=\linewidth]{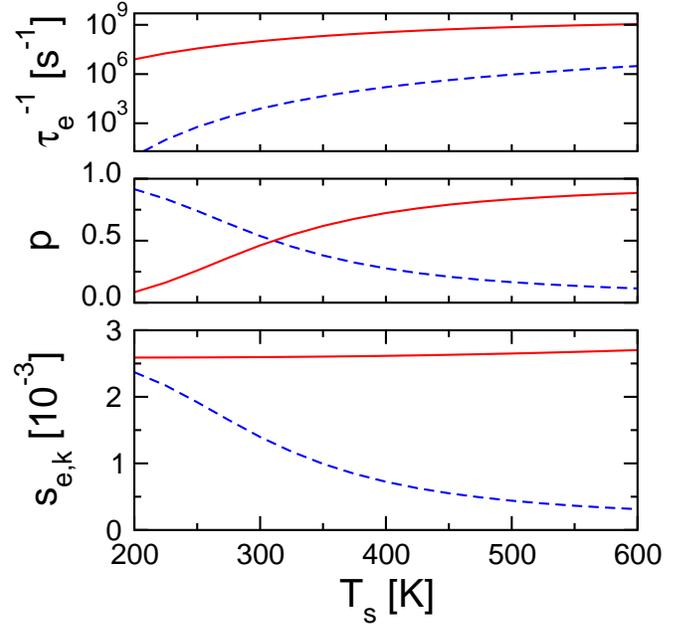}
\caption{The three panels show, as a function of the surface temperature, the quantities of Fig.~\ref{figure4} 
for $T_D=2500{\rm K}$, that is, graphite, and $E_e=0.09{\rm eV}$.
}
\label{figure5}
\end{figure}

The weak coupling between the lowest two bound states in a two-phonon deep surface potential leads to a 
relaxation bottleneck for the electron if it is initially trapped in one of the upper states. In Figs.~\ref{figure4}
and \ref{figure5} we analyze the relaxation bottleneck in more detail as a function of the Debye 
temperature $T_D$ (to realize different potential depths) and the surface temperature $T_s$. The upper panel shows 
the desorption time from the lowest bound state, that is, the desorption time for an electron capable to fall to 
the lowest bound state, and the desorption time from the upper bound states, that is, the desorption time for an 
electron not capable to fall to the lowest bound state. The probability for the electron initially trapped in the 
upper bound states to fall down to the lowest bound states and the probability to desorb to the continuum without 
ever passing through the lowest bound state are shown in the middle panel and the lower panel shows the prompt 
and the kinetic sticking coefficient.

Before we discuss Figs.~\ref{figure4} and~\ref{figure5}, we say a few words about the way we calculated
the quantities shown in the upper and middle panels. The desorption time from the lowest bound state is 
the desorption time for an electron equilibrated with the surface, the quantity we calculated in I, because 
the quasi-stationary occupancy and the equilibrium occupancy coincide and both reside moreover, for the 
considered surface temperatures, mainly on the lowest level. The desorption time from the upper bound 
states we simply calculated from the rate equation (\ref{homorateeqn}) with the lowest bound state excluded.

The probabilities shown in the middle panels we obtained from the following consideration. 
Whether an electron trapped in the upper bound states passes through to the lowest bound state or not 
depends on how large the transition probabilities from the upper bound states to the lowest bound state 
are in comparison to the transition probabilities to the continuum. Hence, the second stage of 
physisorption, that is, the time evolution of the occupancy after the initial trapping, can be 
captured by a rate equation for the occupancy of the upper bound states (\(n=2,3,\dots\)), similar 
to Eq. (\ref{homorateeqn}), but with a loss term to both the continuum and the lowest bound state,
\begin{align}
\frac{\mathrm{d}}{\mathrm{d}t}n_n=&\sum_n \left[W_{n n^\prime} n_{n^\prime}(t)- W_{n^\prime n} n_n(t) \right] \nonumber \\
& -\sum_k W_{k n} n_n(t) -W_{1 n}n_n(t) \nonumber \\
=&D_{n n^\prime} n_{n^\prime}(t)~, \label{upperrelax}
\end{align}
where \(n\) and \(n^\prime\) run over the upper image states. 
Solving Eq.  (\ref{upperrelax}) with the initial condition 
\begin{eqnarray}
n_l(0)=\frac{\sum_k W_{lk}j_k}{\sum_{l,k}W_{lk}j_k} \text{ ,}
\end{eqnarray}
which is the (conditional) probability that the electron is trapped in the \(l^\mathrm{th}\) image state 
under the condition that it is trapped in any of the bound states, we deduce
for the probability for an electron trapped in one of the upper bound states to fall either to the lowest 
bound state (\(f=1\)) or to desorb without falling to the lowest bound state (\(f=c\)), 
\begin{align}
p_f=n_f(t\rightarrow \infty)=\sum_{n,\kappa} W_{fn} \frac{1}{\lambda_\kappa} d_n^{(\kappa)} \sum_l \tilde{d}_l^{(\kappa)} n_l(0) \text{ ,}
\end{align}
where,  \(d^{(\kappa)}_n\) and 
\(\tilde{d}^{(\kappa)}_n\) are the components of the right and left eigenvectors of the matrix \({\bf D}\). 

We now turn our attention to Fig.~\ref{figure4}. It shows the effect of different potential depths realized by 
tuning the Debye temperature. For a shallow potential ($T_D>4029 {\rm K}$) desorption from the lowest level is mainly 
due to direct one-phonon transitions to the continuum; the same type of transition emptying the upper bound states. 
Hence, the desorption time from the lowest bound state and the upper bound states, respectively, differ not too much 
for shallow potentials. For one-phonon deep potentials (\(2707{\rm K} < T_D < 4029{\rm K}\)), however, the cascade of 
two one-phonon processes via the second level yields much larger desorption times from the lowest level compared to 
the desorption time form the upper levels. For a two-phonon deep potential (\(T_D<2707{\rm K}\)), finally, 
the first leg of the cascade, the transition to the second bound state, is a two-phonon transition, which increases 
the desorption time compared to a one-phonon deep potential.

The second level is the link between the upper bound states and the lowest bound state. The ratio of the 
transition probabilities from the second bound state to the lowest bound state, \(W_{12}\), and from the second 
bound state to the continuum, \(W_{c2}\), determines if the electron trickles through after initial trapping
or not, that is, whether it thermalizes with the surface or not. For a one-phonon deep potential both \(W_{12}\) 
and \(W_{c2}\) are due to one-phonon processes, in this case \(W_{12} > W_{c2}\). For a two-phonon deep potential, 
however, the transition from the second to the lowest bound state is enabled by a two-phonon process only. In 
this case, and for moderate surface temperatures, \(W_{12} < W_{c2}\), so that the electron is more likely to 
desorb before relaxing 
to the lowest bound state. As the kinetic sticking coefficient gives the probability of the trapped electron to 
relax to the quasi-stationary occupancy, the drop in the probability for reaching the lowest level at $T_D=2707{\rm K}$, 
which is the one-phonon/two-phonon threshold, translates into the abrupt reduction of the kinetic sticking 
coefficient at $T_D=2707{\rm K}$ (see middle and lower panels of Fig.~\ref{figure4}).

Figure~\ref{figure5} shows the quantities of Fig.~\ref{figure4} as a function of the surface temperature. The
Debye temperature is fixed to the value for graphite. At low surface temperatures the kinetic sticking coefficient 
is only slightly smaller than the prompt sticking coefficient, yet for high surface temperatures their difference 
increases significantly as a consequence of the inhibited thermalization. This can be understood as follows: 
The transition from the second to the first bound state entails the emission of two phonons and the transition
from the second bound state to the continuum requires only the absorption of a single phonon. At low enough
surface temperatures it is however nevertheless possible that the electron drops to the lowest bound state 
because the likelihood of phonon emission is proportional to \(1+n_B\) whilst the likelihood of phonon absorption 
is proportional to \(n_B\). Hence, for sufficiently low surface temperatures, \(W_{12} > W_{c2}\), even when 
$W_{12}$ entails a two-phonon and $W_{c2}$ a one-phonon process, so that the electron has a good chance to 
trickle through. Increasing the surface temperature benefits however \(W_{c2}\) more than \(W_{12}\) so that
\(W_{12}<W_{c2}\), prohibiting the trickling through and leading to a considerable reduction of the kinetic
sticking coefficient at high surface temperatures.

\begin{figure}
\includegraphics[width=\linewidth]{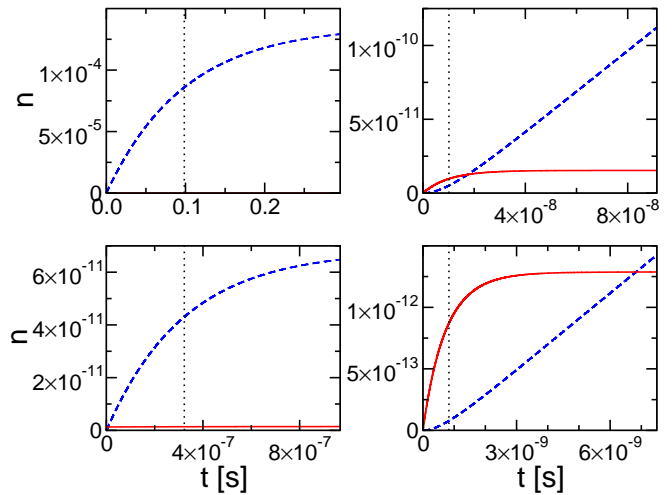}
\caption{Bound state occupancy of the lowest bound state (dashed blue line) and the upper bound
states (solid red line) as a function of time for an unit flux of a Boltzmannian electron with \(k_B T_e=0.1{\rm eV}\).
The left and right panels show the two occupancies on two timescales. The left panel on the scale
of the desorption time  \(\lambda_0^{-1}=\tau_e\) (vertical dashed line in the left panels) and the
right panel on the scale set by \(\lambda_1^{-1}\) (vertical dashed line in the right panels). The upper
two panels are for \(T_s=200{\rm K}\) and \(T_D=2500{\rm K}\) whereas the lower two panels show results for 
\(T_s=600{\rm K}\) and \(T_D=2500{\rm K}\).}
\label{figure6}
\end{figure}

From the discussion of Figs.~\ref{figure4} and \ref{figure5} we conclude that a pronounced relaxation bottleneck 
inhibiting thermalization can only occur for at least two-phonon deep potentials and sufficiently high 
surface temperatures. 

The question arises then on what timescale does the relaxation bottleneck affect
physisorption. To answer this question we analyze, respectively, the time evolution of the occupancy of 
the lowest level and the occupancy of the upper levels of the surface potential under the assumption that 
initially all bound states were empty and that for $t>0$ a stationary unit flux of a Boltzmannian electron 
fills the levels. Accordingly, the occupancy of the lowest state ($n=1$) and the upper states ($n\ge 2$) 
can be determined from Eq.~(\ref{ntotal}) setting \(j_k(t)=0\) for \(t<0\) and \(j_k(t)=j_k\sim k e^{-\beta_e E_k}\) 
for \(t \geq 0\) with \(\sum_k j_k = 1\).

The results of this calculation are shown in Fig.~\ref{figure6} for low (upper two panels) and 
high (lower two panels) surface temperature. Clearly, for times of the order of the desorption time, 
$\tau_e=\lambda_0^{-1}$, indicated by the vertical dashed line in the left panels, the upper levels
are basically empty indicating that a thermalized electron desorbs; for \(T_D=2500{\rm K}\) and 
\(T_s=500{\rm K}\) the quasi-stationary occupancy deviates from the equilibrium 
occupancy less than 3\%. The upper levels are more 
populated than the lower one only for very short timescales, set by \(\lambda_1^{-1}\), indicated 
by the vertical dashed line in the right panels. Since \(\lambda_1^{-1}\ll \lambda_0^{-1}\), 
the relaxation bottleneck does not affect desorption, which still occurs from the equilibrium 
occupancy. It does thus only affect the kinetic sticking coefficient which is significantly smaller 
than the prompt one and actually the one to be used to characterize polarization-induced trapping of 
an electron at a dielectric surface. The relaxation bottleneck is absent in neutral physisorption 
systems because the level spacing is small compared to the Debye energy. Prompt and kinetic sticking 
coefficients are thus almost identical as has been indeed found for neon atoms physisorbing on 
a copper substrate.~\cite{BGB93}

\begin{figure}
\includegraphics[width=\linewidth]{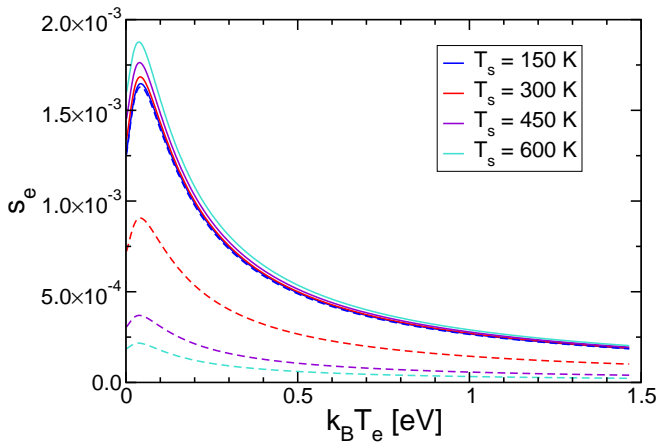}
\caption{Prompt (full line) and kinetic (dashed line) energy-averaged sticking coefficient
for graphite (\(T_D = 2500{\rm K} \)) as a function of the mean energy of the electron and
the surface temperature.}
\label{figure7}
\end{figure}

Figure~\ref{figure7} finally shows for graphite the energy-averaged prompt and kinetic sticking 
coefficients as a function of the mean energy of the incident electron and the surface temperature. As 
two-phonon processes contribute little to the initial trapping of the electron, their most important role 
is to control relaxation to the lowest bound state. In agreement with the foregoing discussion, the kinetic 
sticking coefficient diminishes therefore for higher surface temperatures whereas the prompt sticking 
coefficient is less sensitive to the surface temperature. From Fig.~\ref{figure7} it can be also seen
that even the prompt sticking coefficient for graphite is only at most of the order of of $10^{-3}$, the 
order we also found in our investigation of electron sticking at metallic surfaces.~\cite{BDF09} It is two 
orders of magnitude smaller than the value obtained from a semiclassical estimate~\cite{UN80} whose 
range of applicability is however hard to grasp. We expect it, at best, to be applicable to very low 
mean electron energies, below $0.0026 {\rm eV}$, and rather high electron binding energies, larger than 
$1 {\rm eV}$.~\cite{BDF09}

\section{Conclusions}

As a preparatory step towards a microscopic understanding of the build-up of surface charges at 
dielectric plasma boundaries, we investigated phonon-mediated temporary trapping of an electron on 
a dielectric surface. In our simple model for the polarization-induced interaction of the electron 
and the dielectric surface, the adsorbed electron occupies the bound states of a recoil-corrected 
image potential. Electron energy relaxation responsible for transitions between the image states 
leading to adsorption and eventually to desorption is due to the coupling to an acoustic bulk phonon.

Dielectrics typically used as plasma boundaries are graphite, silicon oxide, aluminium oxide, and bismuth
silicon oxide. They all have large energy gaps blocking internal electronic degrees of freedom and 
small Debye energies compared to the energy difference of at least the lowest two bound surface states. 
Electron physisorption at these boundaries is thus driven by multi-phonon processes.
As in I we presented results for a two-phonon deep surface potential, as it is applicable to graphite, 
where the energy difference between the lowest two bound states is between one and two Debye energies. 
Classifying two-phonon processes by the energy difference they allow to bridge, we included two-phonon 
transition probabilities only for transitions not already triggered by one-phonon processes. Besides
the Debye temperature, which we varied to realize different potential depths, the material parameters
used in the numerical calculations are the ones for graphite.

Similar to physisorption of a neutral particle, sticking and desorption of an electron can be subdivided 
into three characteristic stages. At first, the electron is trapped in one of the upper bound states of the
surface potential. Then the bound state occupancy relaxes to a quasi-stationary occupancy. Finally, 
over the timescale set by the desorption time, the electron desorbs. In order to account for both 
initial trapping and subsequent relaxation we employed a quantum-kinetic rate equation for the occupancy
of the image states. Apart from calculating the energy-resolved and energy-averaged prompt and kinetic 
electron sticking coefficients, which typically turn out to be of the order of $10^{-3}$, we also 
investigated the relative importance of one- and two-phonon processes for the two stages of the sticking 
process. 

The initial trapping is almost entirely due to one-phonon transitions from the continuum, two-phonon processes 
from higher-lying continuum states contribute very little. The relaxation of the bound state occupancy after 
the initial trapping depends strongly on the ratio of the probabilities for downwards transitions to 
the lowest state and upwards transitions to the continuum. For graphite, with its two-phonon deep surface
potential, the upper bound states are linked to the lowest bound state only by a two-phonon process. 
The trapped electron has thus only a slim chance to drop to the lowest bound state, particularly at high 
surface temperatures, which favor transitions back to the continuum. The decreased accessibility of the lowest 
surface state leads to a significant reduction of the  kinetic sticking coefficient compared to the prompt 
sticking coefficient. For the other dielectrics typically used as plasma boundaries, silicon dioxide, 
aluminium oxide, and bismuth silicon oxide, the surface potentials are much deeper because the Debye energy 
for these materials is very small. Hence, more than two phonons are required to link the upper image states 
to the lowest one, the accessibility of the lowest image state is thus even more suppressed, and the 
kinetic sticking coefficient should be accordingly small.

{\it Acknowledgments.}
This work was supported by the Deutsche Forschungsgemeinschaft through the transregional collaborative 
research center TRR 24. F.X.B. and H.F. acknowledge discussions with H. Deutsch in the early stages of 
this investigation.

\end{document}